\documentclass[useAMS,usenatbib]{mn2e}
\usepackage{amsmath}
\usepackage{txfonts}
\usepackage{graphics}
\usepackage{times}

\renewcommand{\vec}[1]{\bmath{#1}}
\newcommand{\mat}[1]{\mathbf{#1}}
\newcommand{\vr}{\vec{r}}
\newcommand{\vq}{\vec{q}}

\newcommand{\vx}{\vec{x}}
\newcommand{\vb}{\vec{b}}
\newcommand{\vg}{\vec{g}}
\newcommand{\vk}{\vec{k}}
\newcommand{\vp}{\vec{p}}
\newcommand{\vs}{\vec{s}}
\newcommand{\vv}{\vec{v}}

\newcommand{\vU}{\vec{U}}
\newcommand{\vPsi}{\vec{\Psi}}
\newcommand{\vDelta}{\vec{\Delta}}

\newcommand{\hq}{\hat{q}}
\newcommand{\hk}{\hat{k}}
\newcommand{\hz}{\hat{z}}
\newcommand{\lam}{\lambda}
\newcommand{\Fp}{\left\langle F' \right\rangle}
\newcommand{\Fpp}{\left\langle F'' \right\rangle}
\DeclareMathOperator{\Tr}{Tr}

\begin{document}


\title[CLPT]{Convolution Lagrangian Perturbation Theory for Biased Tracers}
\author[Carlson et al.]{
    Jordan Carlson$^{1}$\thanks{Email: jwgcarlson@berkeley.edu},
    Beth Reid$^{2}$, and
    Martin White$^{1,2}$
    \\
    $^{1}$ Department of Physics, University of California,
    Berkeley, CA 94720, USA \\
    $^{2}$ Lawrence Berkeley National Laboratory, 1 Cyclotron Road,
    Berkeley, CA 94720, USA
}
\date{\today}
\pagerange{\pageref{firstpage}--\pageref{lastpage}}

\maketitle

\label{firstpage}

\begin{abstract}
We present a new formulation of Lagrangian perturbation theory which
allows accurate predictions of the real- and redshift-space correlation
functions of the mass field and dark matter halos.
Our formulation involves a non-perturbative resummation of Lagrangian
perturbation theory and indeed can be viewed as a partial resummation
of the formalism of \citet{Mat08a,Mat08b} in which we keep exponentiated
all of the terms which tend to a constant at large separation.
One of the key features of our method is that we naturally recover the
Zel'dovich approximation as the lowest order of our expansion for the
matter correlation function.
We compare our results against a suite of N-body simulations and obtain
good agreement for the correlation functions in real-space and for the
monopole correlation function in redshift space.  The agreement becomes
worse for higher multipole moments of the redshift-space, halo correlation
function.
Our formalism naturally includes non-linear bias and explains the strong
bias-dependence of the multipole moments of the redshift-space correlation
function seen in N-body simulations.
\end{abstract}

\begin{keywords}
    gravitation;
    galaxies: haloes;
    galaxies: statistics;
    cosmological parameters;
    large-scale structure of Universe
\end{keywords}


\section{Introduction}
\label{sec:intro}
The observed large-scale structure (LSS) of the universe is a pillar of
modern observational cosmology, providing a window into the primordial
fluctuations, expansion history, and growth rate of perturbations, as well
as allowing tests of the theory of gravity on the largest accessible scales.
The two-point correlation function (or its Fourier transform) is a useful
and relatively simple compression of the cosmological information of interest.
However, the
interpretation of LSS statistics is hampered by two primary uncertainties: LSS
tracers (e.g., galaxies, Ly-$\alpha$ forest, 21 cm) are {\em biased} relative to
the underlying matter density field and are observed in redshift space.  On very
large scales these two effects are simple linear transformations of the
underlying matter density field \citep{Kai87,Efs88}, while on Mpc scales, the
dynamics are highly non-linear, and N-body simulations seem to be required for
quantitative accuracy.  However, on intermediate or quasi-linear scales there is
hope that observable quantities for biased tracers may be accurately modeled
semi-analytically by extending perturbation theory beyond linear order
\citep[see,][for recent work in this direction]{TarNisSai10,ReiWhi11,
OkaTarMat11,Eli11,MPTbreeze,Vlah12}.
\citet{Mat08a,Mat08b} introduced a new perturbative scheme (which we shall
refer to as Lagrangian Resummation Theory; LRT) which addresses both
non-linear biasing and redshift space distortions in a unified framework based
on Lagrangian perturbation theory and that substantially improves upon standard
perturbation theory for the description of both matter and dark matter halo
clustering in the quasilinear regime
\citep{PadWhi09,NohWhiPad09,ReiWhi11,SatoMat11,Ram12b}.
In this paper propose a new resummation scheme which extends the work of
\citet{Mat08a,Mat08b} and results in a more accurate expression for the
two-point correlation function in both real- and redshift-space for both
matter and dark matter halos.

There are several advantages to adopting a Lagrangian description of the LSS.
The well-known Zel'dovich approximation \citep{Zel70} provides a one time-step,
reasonably accurate approximation to the non-linear density field by displacing
Lagrangian particles by the linear theory displacement field
\citep[for a recent examination, see][]{TasZal12a,TasZal12b}.
A distinctive feature of the present work is that we recover the Zel'dovich
result exactly as a limit of our expression for the correlation function of
matter \citep[which is not the case in a similar study by][]{MccSza12}.
The clustering of dark matter halos is of greater interest for the
interpretation of galaxy redshift surveys, since modern galaxy formation
models assume that galaxies form and reside in the gravitational potential
wells of dark matter halos.
Again, the LRT approach is advantageous since a local Lagrangian biasing
scheme provides a better description of the biasing of dark matter halos
in N-body simulations compared with local Eulerian bias
\citep{RotPor11,Bal12,Chan12} and can be extended to include a continuous
galaxy formation history \citep{WanSza12}.

The outline of the paper is as follows.
We begin with a brief review of Lagrangian perturbation theory
in Section \ref{sec:review}.
In Section \ref{sec:zeldovich}, we present our results within the context of
    the Zel'dovich approximation, in which many of the main physical
    effects are present but the algebra is simplified.
In Section \ref{sec:oneloop}, we extend our results to the next higher order
   in perturbation theory, with most of the details and formulae being
   relegated to Appendices.
In Section \ref{sec:results} we present a detailed comparison of our analytic
    theory with high-precision N-body simulations.
We conclude in Section \ref{sec:discuss}.

For plots and numerical comparisons we assume a $\Lambda$CDM cosmology with
$\Omega_m = 0.274$, $\Omega_\Lambda = 0.726$, $h = 0.7$, $n = 0.95$, and
$\sigma_8 = 0.8$.  Our simulation data are derived from a suite of 20 N-body
simulations run with the TreePM code described in \citet{TreePM}.
Each simulation employed $1500^3$ equal mass
($m_p\simeq 7.6\times10^{10}\,h^{-1}M_\odot$)
particles in a periodic cube of side length $1.5\,h^{-1}$Gpc
as described in \citep{ReiWhi11,Whi11}.

\section{Background and review}
\label{sec:review}

In this section we provide a brief review of cosmological perturbation theory,
focusing on the Lagrangian formulation\footnote{See \citet{Ber02} for a
comprehensive (though somewhat dated) review of Eulerian perturbation theory.}
\citep{Buc89,Mou91,Hiv95,TayHam96}.
This material should be sufficient to remind the reader of some essential
terminology, and to establish our notational conventions.
Our discussion is largely drawn from \citet{Mat08a,Mat08b} to which we refer
the reader for further details.

\subsection{Basic definitions}

Cosmological perturbation theory concerns itself with predicting the clustering
properties of cosmological fluids.  In the context of large-scale structure,
(i.e., at late times when baryons and photons have completely decoupled), the
only relevant fluid is the matter fluid, which, on all but the very smallest
scales, interacts only through self-gravitational coupling.

The matter fluid is idealized as a single-streaming, pressureless dust,
characterized at any time $t$ by its mass density $\rho(\vx,t)$ and peculiar
velocity field $\vv(\vx,t)$.  Following common convention, we let $\vx$ denote
position in comoving coordinates, and $t$ denote proper time for a comoving
observer.  The mean value $\bar{\rho}(t) = \langle \rho(\vx,t) \rangle$ of the
mass density decreases as the universe expands like $\bar{\rho} \propto
a^{-3}$, where $a(t)$ is the cosmic scale factor.  (Here and hereafter, we drop
the explicit time dependence in equations where all quantities are to be
evaluated at the same time.)  Deviations from homogeneity are expressed in
terms of the density contrast $\delta(\vx,t)$, defined by the relation
$\rho(\vx) = \bar{\rho}[1 + \delta(\vx)]$.

The most important statistical quantities that can be formed from these fields
are the 2-point correlation function,
\begin{equation}
    \xi(\vr) = \langle \delta(\vx) \delta(\vx+\vr) \rangle ,
\end{equation}
and its Fourier transform, the power spectrum $P(\vk)$, defined by
\begin{equation}
    \langle \tilde{\delta}(\vk) \tilde{\delta}(\vk') \rangle = (2\pi)^3 \delta_D(\vk+\vk') P(\vk) .
\end{equation}
Here $\delta_D$ denotes the 3-dimensional Dirac delta function, and we use the
Fourier transform convention
\begin{equation}
    F(\vx) = \int \frac{d^3k}{(2\pi)^3}~ \tilde{F}(\vk) e^{i\vk\cdot\vx} .
\end{equation}
Angle brackets around a cosmological field, e.g.\ $\langle F \rangle$, signify
an ensemble average of that quantity over all possible realizations of our
universe; in most cases of interest, ergodicity allows us to replace these
ensemble averages with spatial averages over a sufficiently large cosmic
volume.

\subsection{Lagrangian perturbation theory}

In the Lagrangian approach to cosmological fluid dynamics, one traces the
trajectory of an individual fluid element through space and time.  For a fluid
element located at position $\vq$ at some initial time $t_0$, its position at
subsequent times can be written in terms of the Lagrangian displacement field
$\vPsi$,
\begin{equation}
\label{eq:LtoE}
    \vx(\vq,t) = \vq + \vPsi(\vq,t) ,
\end{equation}
where $\vPsi(\vq,t_0) = 0$.  Every element of the fluid is uniquely labeled by
its Lagrangian coordinate $\vq$, so that for a fixed $t$ we may view the mapping
$\vq \leftrightarrow \vx$ as a simple change of variable.

The displacement field $\vPsi(\vq,t)$ fully specifies the motion of the
cosmological fluid.
The aim of Lagrangian Perturbation Theory (LPT) is to find a perturbative
solution for the displacement field,
\begin{equation}
    \label{eq:psin}
    \vPsi(\vq,t) = \vPsi^{(1)}(\vq,t) + \vPsi^{(2)}(\vq,t) + \vPsi^{(3)}(\vq,t) + \cdots .
\end{equation}
The first order solution is the well-known Zel'dovich approximation \citep{Zel70}.
Explicit solutions are known up to fourth order \citep{Ram12a}.


The continuity equation
\begin{equation}
    \label{eq:cont1}
    [1 + \delta(\vx,t)] d^3x = [1 + \delta_0(\vq)] d^3q
\end{equation}
expresses the fact that, for a smoothly evolving fluid, an element $d^3q$
centered at $\vq$ at time $t_0$ is transformed into an element $d^3x$ centered
at $\vx(\vq,t)$ at time $t$.  The initial time $t_0$ may be taken to be early
enough that the initial matter fluctuations $\delta_0(\vq)$ are arbitrarily
small, so that we may formally express the Eulerian density field in terms of
the Lagrangian displacement field as
\begin{equation}
    \label{eq:cont2}
    1 + \delta(\vx,t) = \int d^3q~ \delta_D[\vx - \vq - \vPsi(\vq,t)] .
\end{equation}

\subsection{Biased tracers}
Although small, the initial density fluctuations $\delta_0(\vq)$ provide the
seeds for subsequent structure formation.  In this paper we restrict ourselves
to a local Lagrangian bias model, which posits that the locations of discrete 
tracers at some late time $t$ are determined by the overdensities in the initial 
matter density field.  
More explicitly, let $\delta_R(\vq)$ denote the matter density contrast at the
initial time $t_0$, smoothed on some scale $R$.  (In the end the value of this
smoothing scale will turn out to be irrelevant, but its use helps ensure that
intermediate quantities are well-behaved.)  Consider a collection of discrete
tracers $X$, where $X$ might denote galaxies of a particular type, or halos of
a particular mass range, etc.  The locations $\{\vx_i\}$ of these tracers at
time $t$ may be identified with particular points $\{\vq_i\}$ in the initial
density field by inverting Eqn.~\ref{eq:LtoE}. 
Our hypothesis is that these initial locations $\{\vq_i\}$ are
drawn from a distribution that is a locally biased function of the smoothed
matter density field, i.e.
\begin{equation}
    \rho_X(\vq) = \bar{\rho}_X F[\delta_R(\vq)] .
\end{equation}
Here $\bar{\rho}_X$ is the mean comoving number density of our tracer $X$ and
the function $F(\delta)$ is called the Lagrangian bias function.  The
perturbations in $\rho_{X}(\vq)$ are $\mathcal{O}(\delta_R(\vq))$ and therefore
also arbitrarily small.  When this 
biasing relation is viewed in Eulerian coordinates, it is non-local:
$\rho_X(\vx)$ is depends on the matter density at points other than $\vx$
\citep[e.g.,][]{Cat98,Mat11}.  The corresponding non-local Eulerian bias terms can be most
easily seen in their contribution to the bispectrum of halos
\citep{Bal12,Chan12}.
\citet{Mat11} extends the formalism from
\citet{Mat08a,Mat08b} that we have adopted here to include non-local Lagrangian
bias as well, e.g., the peaks bias model \citep{BBKS}, but we do not explore
those extensions here.
\subsection{Redshift space}

While analyzing the clustering of biased tracers is difficult enough, for
modern surveys we must also deal with the complication of redshift space
distortions.  The position of an object, located at true comoving position
$\vx$, will be mis-identified due to its peculiar velocity along the
line-of-sight, as
\begin{equation}
    \vs = \vx + \frac{\hat{z} \cdot \vv(\vx)}{aH} \hat{z} .
\end{equation}
In this work we adopt the standard ``plane-parallel'' or ``distant-observer''
approximation, in which the line-of-sight direction to each object is taken to
be the fixed direction $\hat{z}$.  While this may seem a poor assumption for
modern wide-area surveys, it has been shown to be sufficient within the level
of current error bars \citep[e.g., Figure 10 of][]{SamPerRac12}.

In the Lagrangian approach, including redshift-space distortions requires only
a simple additive offset of the displacement field.
The peculiar velocity of a fluid element, labeled by its Lagrangian coordinate
$\vq$, is at any time given by
\begin{equation}
    \vv(\vq) = a \dot{\vx}(\vq) = a \dot{\vPsi}(\vq) .
\end{equation}
So in redshift space, the apparent displacement of the fluid element is
\begin{equation}
    \vPsi^s = \vPsi + \frac{\hat{z} \cdot \dot{\vPsi}}{H} \hat{z} .
\end{equation}
To a good approximation the time dependence of the $n$th order term
in Eq.~(\ref{eq:psin}) is given by $\vPsi^{(n)} \propto D^n$.  Therefore
$\dot{\vPsi}^{(n)} = nHf \vPsi^{(n)}$, where $f = d\log D/d\log a$ is the
growth rate, often approximated as $f \approx \Omega_m^{0.6}$.  The mapping to
redshift space may then be achieved, order-by-order, via the matrix
\begin{equation}
    R^{(n)}_{ij} = \delta_{ij} + n f \hat{z}_i \hat{z}_j ,
\end{equation}
as $\vPsi^{s(n)} = R^{(n)} \vPsi^{(n)}$.

\section{Zel'dovich approximation}
\label{sec:zeldovich}

In this section we present a derivation of our new result in the simplified
setting of the Zel'dovich approximation \citep[see also][]{BonCou88}. 
This allows us to sketch the main idea of our approach while avoiding many
of the complications inherent in perturbative calculations.
Several key points regarding the form of the solution are made along the way.

Our starting point is the continuity equation
\begin{equation}
    \left[1 + \delta_X(\vx,t)\right] d^3x = \left[1 + \delta_X(\vq,t_0)\right] d^3q ,
\end{equation}
expressing the conservation of number density for the tracer $X$ between times
$t_0$ and $t$.  Invoking the hypothesis of local Lagrangian biasing, the
quantity on the right-hand side is
\begin{equation}
    1 + \delta_X(\vq,t_0) = F[\delta_R(\vq)] ,
\end{equation}
so that
\begin{eqnarray}
    1 + \delta_X(\vx,t)
    &=& \left|\frac{\partial \vx}{\partial \vq}\right|^{-1} F[\delta_R(\vq)] \\
    &=&\int d^3q~ F[\delta_R(\vq)] \delta_D\left[\vx - \vq - \vPsi(\vq,t)\right]
    \ .
\end{eqnarray}
In the following, we will suppress the explicit dependence on $t$ when there is
no risk of ambiguity.  We now replace the delta function with its Fourier
representation, and also introduce the Fourier transform $\tilde{F}(\lam)$ of
$F(\delta)$,
\begin{align}
  1 + \delta_X(\vx)
  &= \int d^3q~ F[\delta_R(\vq)] \int \frac{d^3k}{(2\pi)^3}
  \ e^{i\vk\cdot[\vx-\vq-\vPsi(\vq)]} \\
  &= \int d^3q \int \frac{d^3k}{(2\pi)^3} \int \frac{d\lam}{2\pi}
  \ \tilde{F}(\lam)
  \ e^{i\left\{\lam \delta_R(\vq) + \vk\cdot[\vx-\vq-\vPsi(\vq)]\right\}} .
\end{align}

The 2-point correlation function
$\xi_X(\vr) = \langle \delta_X(\vx_1) \delta_X(\vx_2) \rangle$
for the biased tracer $X$ is then given by
\begin{eqnarray}
    1 + \xi_X(\vr)
        &=& \int d^3q_1\ d^3q_2
            \int \frac{d^3k_1}{(2\pi)^3} \frac{d^3k_2}{(2\pi)^3}
        \ e^{i\vk_1\cdot(\vx_1-\vq_1)} e^{i\vk_2\cdot(\vx_2-\vq_2)}
        \nonumber \\
   &\times& \int \frac{d\lam_1}{2\pi} \frac{d\lam_2}{2\pi}
   \ \tilde{F}(\lam_1) \tilde{F}(\lam_2)
   \left\langle e^{i[\lam_1 \delta_1 + \lam_2 \delta_2
   - \vk_1\cdot\vPsi_1 - \vk_2\cdot\vPsi_2]} \right\rangle
   \quad ,
\end{eqnarray}
where $\delta_a \equiv \delta_R(\vq_a)$, $\vPsi_a \equiv \vPsi(\vq_a)$, and
$\vr = \vx_2 - \vx_1$.  By statistical homogeneity, the expectation value above
depends only on the difference in Lagrangian coordinates,
$\vq = \vq_2 - \vq_1$.
The change of variables
$\{\vq_1,\vq_2\} \to \left\{\vq,\vec{Q} = (\vq_1+\vq_2)/2\right\}$
then leads to
\begin{equation}
    \label{eq:1+xi}
    1 + \xi_X(\vr)
        = \int d^3q
            \int \frac{d^3k}{(2\pi)^3} e^{i\vk\cdot(\vq-\vr)}
            \int \frac{d\lam_1}{2\pi} \frac{d\lam_2}{2\pi}
            \ \tilde{F}_1 \tilde{F}_2\,K(\vq,\vk,\lam_1,\lam_2) ,
\end{equation}
where we have defined
\begin{equation}
    \label{eq:K}
    K(\vq,\vk,\lam_1,\lam_2) = \left\langle e^{i(\lam_1 \delta_1 + \lam_2 \delta_2 + \vk\cdot\vDelta)} \right\rangle ,
\end{equation}
and $\Delta \equiv \vPsi_2 - \vPsi_1$.  This expression is the exact
configuration space analog of Eq. (9) in \citet{Mat08b}.

The cumulant expansion theorem allows us to expand the expectation value
in Eq.~(\ref{eq:K}) in terms of cumulants,
\begin{equation}
    \langle e^{iX} \rangle = \exp\left[ \sum_{N=1}^\infty \frac{i^N}{N!} \langle X^N \rangle_c \right] ,
\end{equation}
where $\langle X^N \rangle_c$ denotes the $N$th cumulant of the random variable
$X$.
The field $\delta_R(\vq)$ is a smoothed version of the linear density field
$\delta_L(\vq)$, and is therefore Gaussian.  Within the Zel'dovich
approximation, the displacement field
\begin{equation}
    \vPsi(\vq) = \int \frac{d^3k}{(2\pi)^3}~ e^{i\vk\cdot\vq} \frac{i\vk}{k^2} \tilde{\delta_L}(\vk) ,
\end{equation}
is linear in $\delta_L$, hence also Gaussian.
Thus, in the applying the cumulant expansion theorem to Eq. \eqref{eq:K},
only the second cumulant survives,
\begin{eqnarray}
  \left\langle \left(\lam_1 \delta_1 + \lam_2 \delta_2
  + \vk\cdot\vDelta\right)^2 \right\rangle_c
  &=& (\lam_1^2 + \lam_2^2) \sigma_R^2 + A_{ij} k_i k_j \nonumber \\
  &+& 2 \lam_1 \lam_2 \xi_R + 2(\lam_1 + \lam_2) U_i k_i ,
\end{eqnarray}
where we have defined
\begin{align}
    \sigma_R^2 &= \langle \delta_1^2 \rangle_c = \langle \delta_2^2 \rangle_c , &
    \xi_R(\vq) &= \langle \delta_1 \delta_2 \rangle_c , \\
    A_{ij}(\vq) &= \langle \Delta_i \Delta_j \rangle_c , &
    U_i(\vq) &= \langle \delta_1 \Delta_i \rangle_c = \langle \delta_2 \Delta_i \rangle_c .
\end{align}
Eq. (\ref{eq:K}) then evaluates to
\begin{equation}
    \label{eq:K_ZA}
    K = \exp\left[-\frac{1}{2} (\lam_1^2 + \lam_2^2) \sigma_R^2 - \frac{1}{2} A_{ij} k_i k_j - \lam_1 \lam_2 \xi_R - (\lam_1 + \lam_2) U_i k_i \right] .
\end{equation}

The quantity $\sigma_R^2$ is simply the variance of the smoothed linear density
field, while $\xi_R(\vq) = \langle \delta_R(\vq_1) \delta_R(\vq_2) \rangle$ is
the corresponding smoothed linear correlation function.  The matrix $A_{ij}$
may be decomposed as
\begin{equation}
  A_{ij}(\vq) = 2\left[\sigma_\eta^2 - \eta_\perp(q)\right] \delta_{ij}
              + 2\left[\eta_\perp(q) - \eta_\parallel(q)\right] \hq_i \hq_j ,
\end{equation}
where $\sigma_\eta^2 \equiv \frac{1}{3} \langle |\vPsi|^2 \rangle$ is the 1-D
dispersion of the displacement field, and $\eta_\parallel$ and $\eta_\perp$ are
the transverse and longitudinal components of the Lagrangian 2-point function,
$\eta_{ij}(\vq) = \left\langle \Psi_i(\vq_1) \Psi_j(\vq_2) \right\rangle$.
The vector $U_i(\vq)=U(q)\,\hq_i$ is the cross-correlation between the linear
density field and the Lagrangian displacement field.
In the Zel'dovich approximation these quantities are given by
\begin{gather}
    \sigma_\eta^2 = \frac{1}{6\pi^2} \int_0^\infty dk~ P_L(k) , \\
    \eta_\perp(q) = \frac{1}{2\pi^2} \int_0^\infty dk~ P_L(k)~ \frac{j_1(kq)}{kq} , \\
    \eta_\parallel(q) = \frac{1}{2\pi^2} \int_0^\infty dk~ P_L(k)~ \left[j_0(kq) - 2 \frac{j_1(kq)}{kq}\right] , \\
    U(q) = -\frac{1}{2\pi^2} \int_0^\infty dk~ k P_L(k)~ j_1(kq) .
\end{gather}
Up to factors of 2 and $f$, these expressions are identical to the Eulerian
velocity correlators in linear theory \citep[e.g.][]{Fis95,ReiWhi11}, which is not
surprising since $\vv_L = f \Psi$ in the Zel'dovich approximation.

\subsection{Exact results for matter}

At this point we pause to consider the unbiased case, where $F(\delta) = 1$ or
$\tilde{F}(\lam) = 2\pi \delta_D(\lam)$.  In this limit Eq. \eqref{eq:1+xi}
reduces to
\begin{align}
    1 + \xi^{(ZA)}(\vr)
        &= \int d^3q \int \frac{d^3k}{(2\pi)^3} e^{i\vk\cdot(\vq-\vr)} e^{-\frac{1}{2} A_{ij} k_i k_j} \\
        &= \int \frac{d^3q}{(2\pi)^{3/2} |A|^{1/2}} e^{-\frac{1}{2} (\vr-\vq)^\top \mat{A}^{-1} (\vr-\vq)} ,
\end{align}
after carrying out the Gaussian integral over $\vk$ analytically.  This is an
exact expression for the real-space matter correlation function within the
Zel'dovich approximation.  It has the apparent form of a Gaussian convolution
kernel, except for the fact that the matrix $A_{ij}$ is a function of $\vq$.
Indeed, we see that $\xi^{(ZA)}$ arises entirely from the scale-dependence of
this Lagrangian correlator.

The smoothing of the acoustic feature is often modeled as a convolution of
the linear correlation function by a Gaussian kernel, with the smoothing scale
estimated at lowest order by $2 \sigma_\eta^2$.
We can massage our expression into a similar form by noting that $A_{ij}$
can be written as the sum
\begin{equation}
    A_{ij}(\vq) = B_{ij} + C_{ij}(\vq) ,
\end{equation}
where $B_{ij} = 2 \sigma_\eta^2 \delta_{ij}$ is scale-independent.
Then, by the same reasoning as is used to show that the convolution of two
Gaussians is a Gaussian, we can write
\begin{equation}
    1 + \xi^{(ZA)}(\vr)
    = \int \frac{d^3q}{(2\pi)^{3/2} |B|^{1/2}}
    \ e^{-\frac{1}{2}(\vr-\vq)^\top \mat{B}^{-1} (\vr-\vq)} [1 + \chi(\vq)] ,
\label{eq:conv}
\end{equation}
where we have defined
\begin{equation}
    1 + \chi(\vq) = \int \frac{d^3p}{(2\pi)^{3/2} |C|^{1/2}}
    \ e^{-\frac{1}{2} (\vq-\vp)^\top \mat{C}^{-1} (\vq-\vp)} .
\end{equation}
Eq.~(\ref{eq:conv}) is a proper Gaussian convolution, since the matrix
$\mat{B}$ is independent of $\vq$.  The quantity $\chi(\vq)$ may therefore be
viewed as an analog of the linear correlation function.
Indeed, the two are quite similar.
These observations provide analytic justification to conventional wisdom,
first pointed out in \citet{Bha96}, that non-linear structure growth causes
a Gaussian smearing of the clustering signal.
In our approach, this result is obtained at leading order, within the
Zel'dovich approximation.

\subsection{Perturbative expansion for biased tracers}
\label{sec:zeldovich-biased}

Returning to the case of biased tracers, consider again Eq. (\ref{eq:K_ZA}).
In the unbiased case the $\vk$ integration in Eq. (\ref{eq:1+xi}) took the form
of a Gaussian integral, which we carried out analytically.  In the biased case,
we can achieve the same thing if we first partially expand Eq. (\ref{eq:K_ZA})
as
\begin{eqnarray}
    K &=& e^{-\frac{1}{2}(\lam_1^2 + \lam_2^2) \sigma_R^2}
    e^{-\frac{1}{2} \vk^T A \vk} \left[1 - \lam_1 \lam_2 \xi_R
    - (\lam_1 + \lam_2) U_i k_i \right.  \nonumber \\
   &+& \frac{1}{2} \lam_1^2 \lam_2^2 \xi_R^2
    +  \frac{1}{2} (\lam_1+\lam_2)^2 U_i U_j k_i k_j \nonumber \\
   &+& \left. \lam_1 \lam_2 (\lam_1+\lam_2) \xi_R U_i k_i + O(P_L^3) \right] .
\label{eq:K.expanded}
\end{eqnarray}
We may justify this choice of expansion by noting that both $\xi_R(\vq)$ and
$U_i(\vq)$ vanish in the large-scale limit $|\vq| \to \infty$, while
$\sigma_R^2$ and $A_{ij}(\vq)$ approach non-zero values.  In the language of
perturbation theory, keeping these terms exponentiated therefore amounts to an
non-perturbative resummation of the dominant large-scale contributions.

\citet{TasZal12a,TasZal12b} have recently emphasized the importance of not
splitting the effects of bulk flows across orders in perturbation theory.
The resummation described above has this property, which is not shared by
the resummations used in LRT or RPT \citep{RPT}.

To get from Eq. (\ref{eq:K.expanded}) to an expression for $\xi_X(\vr)$, we
must integrate $K$ over $\lam_1, \lam_2$, $\vk$, and $\vq$.  The $\lam_1$ and
$\lam_2$ integrations may be evaluated via the identity \citep{Mat08b}
\begin{equation}
    \int \frac{d\lam}{2\pi}~ \tilde{F}(\lam)~ (i\lam)^n~ e^{-\frac{1}{2} \lam^2 \sigma_R^2}
        = \int \frac{d\delta}{\sqrt{2\pi} \sigma_R}~ e^{-\delta^2/2\sigma_R^2} \frac{d^nF}{d\delta^n}
        \equiv \left\langle F^{(n)} \right\rangle ,
\end{equation}
where $\left\langle F^{(n)} \right\rangle$ is the expectation value of the
$n$th derivative of the Lagrangian bias function $F(\delta)$
(see Appendix \ref{sec:bias} for details).
Application of this identity leads to
\begin{eqnarray}
  L(\vq,\vk) &\equiv&
  \int \frac{d\lam_1}{2\pi} \frac{d\lam_2}{2\pi}
  \ \tilde{F}(\lam_1)\tilde{F}(\lam_2)\,K(\vq,\vk,\lam_1,\lam_2) \\
  &=& e^{-\frac{1}{2} A_{ij} k_i k_j} \left[
  1 + \Fp^2 \xi_R + 2i \Fp U_i k_i + \frac{1}{2} \Fpp^2 \xi_R^2 \right.
  \nonumber \\
  &-& (\Fpp + \Fp^2) U_i U_j k_i k_j + 2i \Fp \Fpp \xi_R U_i k_i
  \nonumber \\
  &+& \left. O(P_L^3) \right] .
  \label{eq:L_ZA}
\end{eqnarray}
The $\vk$ integration reduces to a series of multi-variate Gaussian integrals
of the form
\begin{equation}
  \int \frac{d^3k}{(2\pi)^3}\ e^{-\frac{1}{2} A_{ij} k_i k_j}
                            \ e^{i\vk\cdot(\vq-\vr)}
                            \ k_{i_1} \cdots k_{i_r} \quad .
\end{equation}
Appendix \ref{sec:reference} reviews the relevant formulae.
In the end we obtain
\begin{eqnarray}
  M(\vr,\vq) &\equiv&
  \int \frac{d^3k}{(2\pi)^3}~ e^{i\vk\cdot(\vq-\vr)} L(\vq,\vk) \\
  &=& \frac{1}{(2\pi)^{3/2} |A|^{1/2}}
  \ e^{-\frac{1}{2} (\vr-\vq)^T \mat{A}^{-1} (\vr-\vq)}
  \left[ 1 + \Fp^2 \xi_R \right. \nonumber \\
  &-& 2 \Fp U_i g_i + \frac{1}{2} \Fpp^2 \xi_R^2
  - (\Fpp + \Fp^2) U_i U_j G_{ij} \nonumber \\
  &-& \left.  2\Fp \Fpp \xi_R U_i g_i + O(P_L^3) \right] ,
\label{eq:M_ZA}
\end{eqnarray}
where
\begin{equation}
    \label{eq:g_i}
    \vg \equiv A^{-1} (\vq - \vr) , \quad
    G_{ij} \equiv (A^{-1})_{ij} - g_i g_j .
\end{equation}
Our final expression for the correlation function is
\begin{equation}
    1 + \xi_X(\vr) = \int d^3q~ M(\vr,\vq) .
\label{eq:Z_ZA}
\end{equation}
The remaining integration over $\vq$ must be performed numerically.

Note well that, although our calculation is very similar to that of
\citet{Mat08b}, our result for $\xi_X(r)$ is \emph{not} simply the Fourier
transform of his Eq.~(34) for $P_\text{obj}(k)$
The difference lies in our choice of expansion in Eq.~(\ref{eq:K.expanded}).
As discussed previously, the matrix $A_{ij}(\vq)$ is the sum of a constant
term $2 \sigma_\eta^2 \delta_{ij}$ and a scale-dependent remainder $C_{ij}(\vq)$.
In \citet{Mat08b} only the constant piece is exponentiated while the rest is
expanded, i.e.
\begin{equation}
  K_{\rm Mat} = e^{-\frac{1}{2}(\lam_1^2 + \lam_2^2) \sigma_R^2}
       e^{-\sigma_\eta^2 k^2}
        \left[1 - \frac{1}{2} \vk^\top \mat{C} \vk - \lam_1 \lam_2 \xi_R
       - (\lam_1 + \lam_2) U_i k_i + \cdots\right] .
\label{eq:K.Matsubara}
\end{equation}
Our approach may be seen as a partial resummation of the result of
\citet{Mat08b}, and as such we expect it to be more accurate on small scales.

Before we leave this section it is worth noting the manner in which the
bias terms enter in Eq.~(\ref{eq:L_ZA}).  In particular note the term which
goes as $\langle F'\rangle\,\langle F''\rangle$ at the end of the $3^{\rm rd}$
line and the $\langle F''\rangle^2$ term at the end of the $2^{\rm nd}$ line.
For highly biased halos, assuming the peak-background split to compute the bias,
$\langle F''\rangle\propto\langle F'\rangle^2\propto b^2$, so these terms can
come in with (apparently) large powers of $b$, beyond the $b^2$ terms which
one would naturally expect in a 2-point function \citep[see also][]{ReiWhi11}.
In our calculation using the Zel'dovich approximation and local Lagrangian
bias we see these important contributions arise from $2^{\rm nd}$ order bias.

\subsection{Redshift space}

Thus far we have concentrated on real space results, however the transition to
redshift space is easily achieved.  Recall that the displacement field in
redshift space is given by $\vPsi^s = \vPsi + H^{-1} (\hz\cdot\dot{\vPsi})
\hz$.  In the Zel'dovich approximation $\vPsi \propto D(t)$, so
\begin{equation}
    \Psi_i^{s(ZA)} = \left(\delta_{ij} + f \hz_i \hz_j\right) \Psi_j^{(ZA)} .
\end{equation}
Our previous derivation remains valid, we need only make the substitutions
\begin{align}
    U_i & \to U_i^s = (\delta_{ij} + f \hz_i \hz_j) U_j , \\
    A_{ij} & \to A_{ij}^s = (\delta_{ik} + f \hz_i \hz_k) (\delta_{jl} + f \hz_j \hz_l) A_{kl} .
\end{align}
This slightly complicates the evaluation of the $\vq$ integration in Eq.
\eqref{eq:Z_ZA}, in that we can no longer use azimuthal symmetry to reduce it
to a 2-D integral.  Nevertheless, the full 3-D integral is still feasible
numerically, and the redshift space correlation function $\xi_X^s(\vs)$ may be
easily calculated.

\subsection{Linear theory limit}
\label{sec:zeldovich-linear}

Standard Eulerian perturbation theory describes an expansion for the power
spectrum of the form,
\begin{equation}
    P(k) = P^{(1)}(k) + P^{(2)}(k) + \cdots
\end{equation}
where $P^{(n)}$ is $O(P_L^n)$.  Unfortunately this expansion does not translate
into a well-defined perturbative expansion for $\xi(r)$, as the Fourier
transform of $P^{(n)}$ diverges for $n > 1$.  Nevertheless, the linear theory
correlation function is well-defined, and our approach should reproduce this
limit when $P_L$ is small.  We now show that this is indeed the case.

In the Zel'dovich approximation, the correlators $A_{ij}$ and $U_i$ are given
by linear integrals of $P_L$,
\begin{equation}
    \label{eq:AU}
    \begin{aligned}
    A_{ij}(\vq) &= \int \frac{d^3k}{(2\pi)^3}~ \left[2 - e^{i\vk\cdot\vq} - e^{-i\vk\cdot\vq}\right] \frac{-k_i k_j}{k^4} P_L(k) , \\
    U_i(\vq) &= \int \frac{d^3k}{(2\pi)^3}~ e^{i\vk\cdot\vq}~ \frac{i k_i}{k^2} P_L(k) .
    \end{aligned}
\end{equation}
The quantity $M(\vr,\vq)$ defined in Eq. \eqref{eq:M_ZA} is therefore
ill-defined in the limit $P_L \to 0$.  To make our discussion precise, we
replace the matrix $A_{ij}$ in this expression by
\begin{equation}
    B_{ij}(\vq) = \beta^2 \delta_{ij} + \epsilon A_{ij}(\vq) ,
\end{equation}
where $\beta$ is a regularizing parameter that will eventually be set to zero,
and $\epsilon$ is a book-keeping parameter to help keep track of powers of
$P_L$.  Thus we write
\begin{eqnarray}
  1 + \xi_X(\vr) &=& \lim_{\beta \to 0}
  \int \frac{d^3q}{(2\pi)^{3/2} |B|^{1/2}}
  \ e^{-\frac{1}{2} (\vr-\vq)^\top \mat{B}^{-1} (\vr-\vq)}
  \left[1 + \right. \nonumber \\
  && \left. 2 \epsilon \Fp \vU^\top \mat{B}^{-1} (\vr-\vq)
     + \epsilon \Fp^2 \xi_L + O(\epsilon^2) \right] \\
  &=& 1 + \epsilon \xi_X^{(1)}(\vr) + \epsilon^2 \xi_X^{(2)}(\vr) + \cdots .
\end{eqnarray}
The linear contribution is then given by
$\xi_X^{(1)} = \left.\partial \xi_X/\partial\epsilon\right|_{\epsilon=0}$.

Using the identities
\begin{gather}
    \frac{\partial}{\partial\epsilon} \det \mat{B} = (\det \mat{B}) \Tr \left[\mat{B}^{-1} \frac{\partial \mat{B}}{\partial\epsilon}\right] , \\
    \frac{\partial \mat{B}^{-1}}{\partial\epsilon} = -\mat{B}^{-1} \frac{\partial \mat{B}}{\partial\epsilon} \mat{B}^{-1} ,
\end{gather}
we have
\begin{eqnarray}
  \frac{\partial \xi_X}{\partial\epsilon} &=&
  \int \frac{d^3q}{(2\pi)^{3/2} |B|^{1/2}}
  \ e^{-\frac{1}{2} (\vr-\vq)^\top \mat{B}^{-1} (\vr-\vq)} \nonumber \\
  &\times&\left[\frac{1}{2} (\vr-\vq)^\top \mat{B}^{-1} \mat{A}
  \mat{B}^{-1} (\vr-\vq) - \frac{1}{2} \Tr (\mat{B}^{-1} \mat{A})
  \right. \nonumber\\
  &+& \left. 2 \Fp \vU^\top \mat{B}^{-1} (\vr-\vq) + \Fp^2 \xi_L
   + O(\epsilon) \right] \\
  &\stackrel{\epsilon \to 0}{=}&
  \int \frac{d^3q}{(2\pi)^{3/2} \beta^3}\ e^{-(\vr-\vq)^2/2\beta^2}
  \left[\frac{1}{2} \beta^{-4} (\vr-\vq)^\top \mat{A} (\vr-\vq) \right.
  \nonumber \\
  &-& \left. \frac{1}{2} \beta^{-2} \Tr \mat{A} +
  2 \Fp \beta^{-2} \vU^\top (\vr-\vq) + \Fp^2 \xi_L \right] .
\end{eqnarray}
Integrating by parts, and noting that
\begin{equation}
    \lim_{\beta \to 0} \frac{1}{(2\pi)^{3/2} \beta^3}~ e^{-(\vr-\vq)^2/2\beta^2} = \delta_D(\vr-\vq) ,
\end{equation}
we obtain
\begin{equation}
    \xi_X^{(1)}(\vr) = \frac{1}{2} \frac{\partial^2 A_{ij}}{\partial r_i \partial r_j}(\vr) - 2 \Fp \frac{\partial U_i}{\partial r_i}(\vr) + \Fp^2 \xi_L(\vr) .
\end{equation}
We see immediately from Eq. \eqref{eq:AU} that
\begin{equation}
    \frac{\partial^2 A_{ij}}{\partial q_i \partial q_j}(\vq) = 2 \xi_L(\vq) ,
    \quad
    \frac{\partial U_i}{\partial q_i}(\vq) = -\xi_L(\vq) .
\end{equation}
Therefore the linear theory limit of our result is
\begin{equation}
    \xi_X^{(1)}(\vr) = \left[1 + \Fp\right]^2 \xi_L(\vr) ,
\end{equation}
in agreement with standard perturbation theory.

\section{Higher order}
\label{sec:oneloop}

We now repeat the derivation of the previous section, this time with the aim of
extending our result to one order beyond the Zel'dovich approximation.
Many of the technical details are relegated to appendices.

We pick up the track following Eq.~\eqref{eq:K}, prior to which we make no use
of the Zel'dovich approximation.  With the help of the multinomial theorem, the
cumulant expansion of Eq. \eqref{eq:K} in the general case can be written
\begin{align}
  \log K &= \sum_{N=1}^\infty \frac{i^N}{N!}
  \left\langle(\lam_1\delta_1+\lam_2\delta_2+\vk\cdot\vDelta)^N
  \right\rangle_c \\
   \label{eq:logK}
   &= \sum_{m,n,r} \frac{i^{m+n+r}}{m!n!r!} \lam_1^m \lam_2^n
   k_{i_1} \dots k_{i_r}
   \left\langle\delta_1^m\delta_2^n\Delta_{i_1}\dots\Delta_{i_r}\right\rangle_c
   .
\end{align}
The cumulants
$\left\langle\delta_1^m\delta_2^n\Delta_{i_1}\dots\Delta_{i_r}\right\rangle_c$
are the key ingredients in our theory.  In the following we refer to them
generally as ``Lagrangian correlators.''  As emphasized previously, they are
functions of $\vq$ only, so their tensor structure places severe restrictions
on their functional form (see Appendix \ref{sec:correlators}).
Moreover, due to the properties of Gaussian random fields, a cumulant of
order $m+n+r$ must be at least of order $m+n+r-1$ in the linear power
spectrum $P_L$ \citep[e.g.][]{Ber02}.
An expansion in cumulant order therefore corresponds to a perturbative
expansion in powers of $P_L$.

For convenience, we assign different symbols to these Lagrangian correlators
based on their tensor rank $r$.
For $r=0$, since $\delta_R$ is Gaussian, the only non-vanishing cumulants are
\begin{equation}
  \langle \delta_1^2 \rangle_c = \langle \delta_2^2 \rangle_c \equiv \sigma_R^2,
  \qquad \langle \delta_1 \delta_2 \rangle_c \equiv \xi_R(\vq) .
\end{equation}
For $r=1$, 2, and 3 we denote
\begin{align}
 U_i^{mn} &\equiv \langle \delta_1^m \delta_2^n \Delta_i \rangle_c , \\
 A_{ij}^{mn}&\equiv\langle \delta_1^m\delta_2^n\Delta_i\Delta_j\rangle_c , \\
 W_{ijk}^{mn}&\equiv\langle\delta_1^m\delta_2^n\Delta_i\Delta_j\Delta_k\rangle_c
 \quad .
\end{align}
Explicit expressions for these quantities may be found in
Appendix \ref{sec:correlators}.
Since they arise frequently, and to remain consistent with the previous
section, we also adopt the shorthand
\begin{equation}
    U_{i}^{10} \to U_i , \quad
    A_{ij}^{00} \to A_{ij} , \quad \text{and }
    W_{ijk}^{00} \to W_{ijk} .
\end{equation}

In this notation, we evaluate Eq. \eqref{eq:logK} up to cumulants of order
three,
\begin{eqnarray}
  \log K &=& -\frac{1}{2}(\lam_1^2 + \lam_2^2) \sigma_R^2
  - \frac{1}{2} A_{ij} k_i k_j - \lam_1 \lam_2 \xi_R \nonumber \\
  &-& (\lam_1 + \lam_2) U_i k_i - \frac{i}{6} W_{ijk} k_i k_j k_k
   - \frac{i}{2} (\lam_1 + \lam_2) A_{ij}^{10} k_i k_j \nonumber \\
  &-&\frac{i}{2} (\lam_1^2 + \lam_2^2) U_{i}^{20} k_i
   - i \lam_1 \lam_2 U_{i}^{11} k_i + O(P_L^3) .
\label{eq:logK.2}
\end{eqnarray}
We recover $K$ by exponentiating.  Of the eight terms in the above expression,
only the first two have non-zero limits as $|\vq| \to \infty$, \emph{and}
include $O(P_L)$ contributions.  As in the Zel'dovich case, we leave these two
terms exponentiated while expanding the rest, thus
\begin{eqnarray}
  K &=& e^{-\frac{1}{2} (\lam_1^2 + \lam_2^2) \sigma_R^2
  - \frac{1}{2} A_{ij} k_i k_j} \bigg[1 - \lam_1 \lam_2 \xi_R -
  (\lam_1 + \lam_2) U_i k_i + \frac{1}{2} \lam_1^2 \lam_2^2 \xi_R^2
  \nonumber \\
  &+& \frac{1}{2} (\lam_1 + \lam_2)^2 U_i U_j k_i k_j +
  \lam_1 \lam_2 (\lam_1 + \lam_2) \xi_R U_i k_i \nonumber \\
  &-& \frac{i}{6} W_{ijk} k_i k_j k_k -
  \frac{i}{2} (\lam_1 + \lam_2) A_{ij}^{10} k_i k_j -
  \frac{i}{2} (\lam_1^2 + \lam_2^2) U_i^{20} k_i \nonumber \\
  &-& i \lam_1 \lam_2 U_i^{11} k_i + O(P_L^3) \bigg] .
\label{eq:K.3}
\end{eqnarray}

As in Section \ref{sec:zeldovich-biased}, we must now integrate with
respect to $\lam_1$, $\lam_2$, $\vk$, and $\vq$.
The analog of Eq.~\eqref{eq:L_ZA} for the one-loop case is
\begin{eqnarray}
  L &=& e^{-\frac{1}{2} A_{ij} k_i k_j}
  \Bigg[1 + \Fp^2 \xi_R + 2i \Fp U_i k_i + \frac{1}{2} \Fpp^2 \xi_R^2
  \nonumber \\
  &-& (\Fpp + \Fp^2) U_i U_j k_i k_j + 2i \Fp \Fpp \xi_R U_i k_i \nonumber \\
  &-& \frac{i}{6} W_{ijk} k_i k_j k_k - \Fp A_{ij}^{10} k_i k_j +
  i \Fpp U_i^{20} k_i \nonumber \\
  &+& i \Fp^2 U_i^{11} k_i + O(P_L^3) \Bigg] ,
\end{eqnarray}
Analogous to Eq. \eqref{eq:M_ZA}, the $\vk$ integration gives (see Appendix
\ref{sec:reference})
\begin{eqnarray}
 M &=& \frac{1}{(2\pi)^{3/2} |A|^{1/2}}
  \ e^{-\frac{1}{2} (\vr-\vq)^T \mat{A}^{-1} (\vr-\vq)}
  \Bigg[1 + \Fp^2 \xi_R \nonumber \\
  &-& 2 \Fp U_i g_i + \frac{1}{2} \Fpp^2 \xi_R
  - [\Fpp + \Fp^2] U_i U_j G_{ij} \nonumber \\
  &-& 2\Fp \Fpp \xi_R U_i g_i 
   + \frac{1}{6} W_{ijk} \Gamma_{ijk} - \Fp A_{ij}^{10} G_{ij}
  \nonumber \\
  &-& \Fpp U_i^{20} g_i - \Fp^2 U_i^{11} g_i + O(P_L^3) \Bigg] .
\label{eq:M_final}
\end{eqnarray}
where $g_i$ and $G_{ij}$ are defined in Eq. \eqref{eq:g_i}, and
\begin{equation}
    \Gamma_{ijk} \equiv (A^{-1})_{ij} g_k + (A^{-1})_{ki} g_j + (A^{-1})_{jk} g_i - g_i g_j g_k ,
\end{equation}
Our final expression for the real-space correlation function $\xi_X(\vr)$ is
given once again by Eq. \eqref{eq:Z_ZA}, with $M(\vr,\vq)$ given by Eq.
\eqref{eq:M_final} up to $O(P_L^2)$.  The redshift-space correlation function
$\xi_X^s(\vs)$ is obtained by replacing the real-space Lagrangian correlators
by their redshift-space counterparts.

\begin{figure}
    \begin{center}
    \resizebox{3in}{!}{\includegraphics{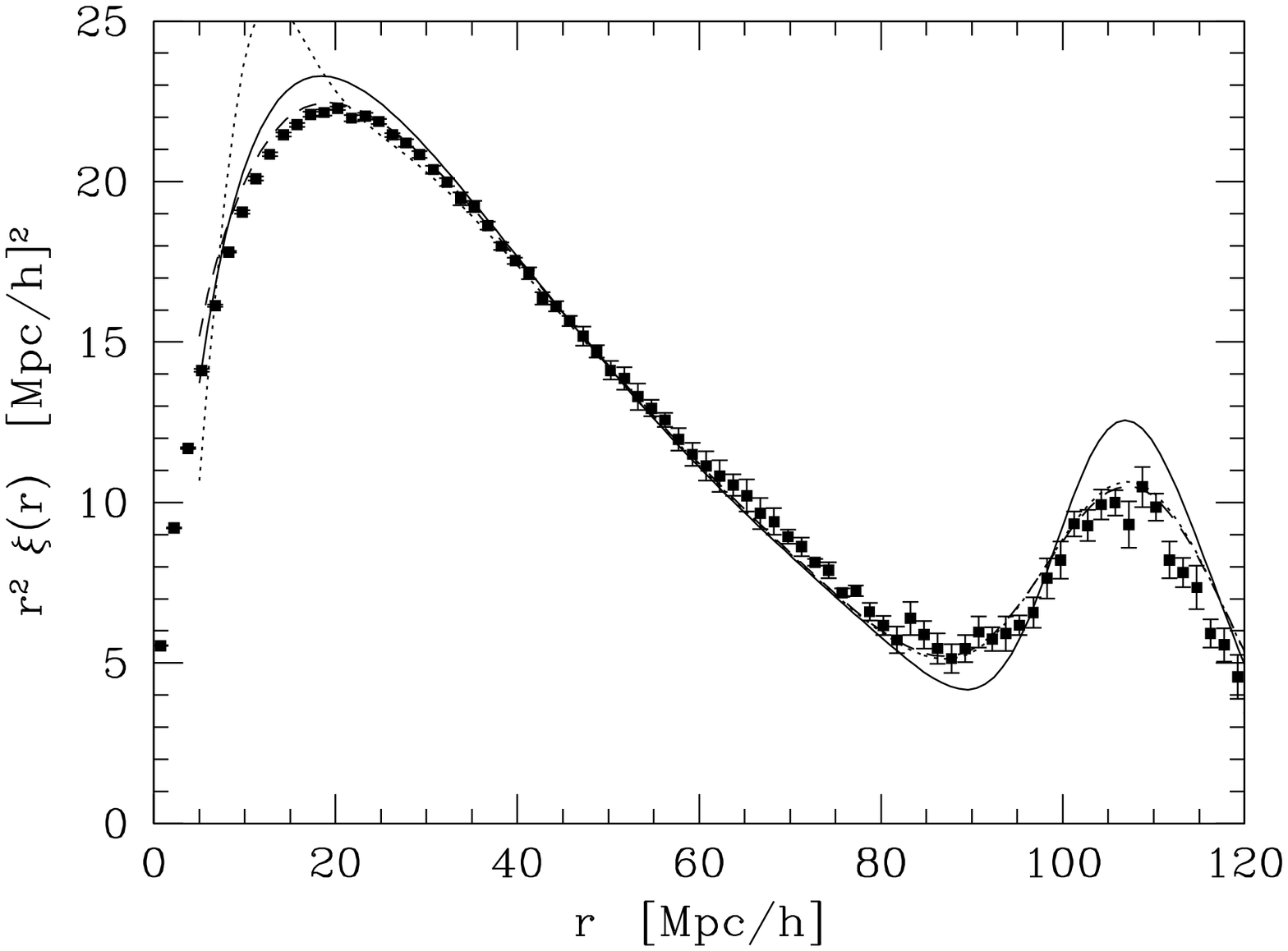}}
    \resizebox{3in}{!}{\includegraphics{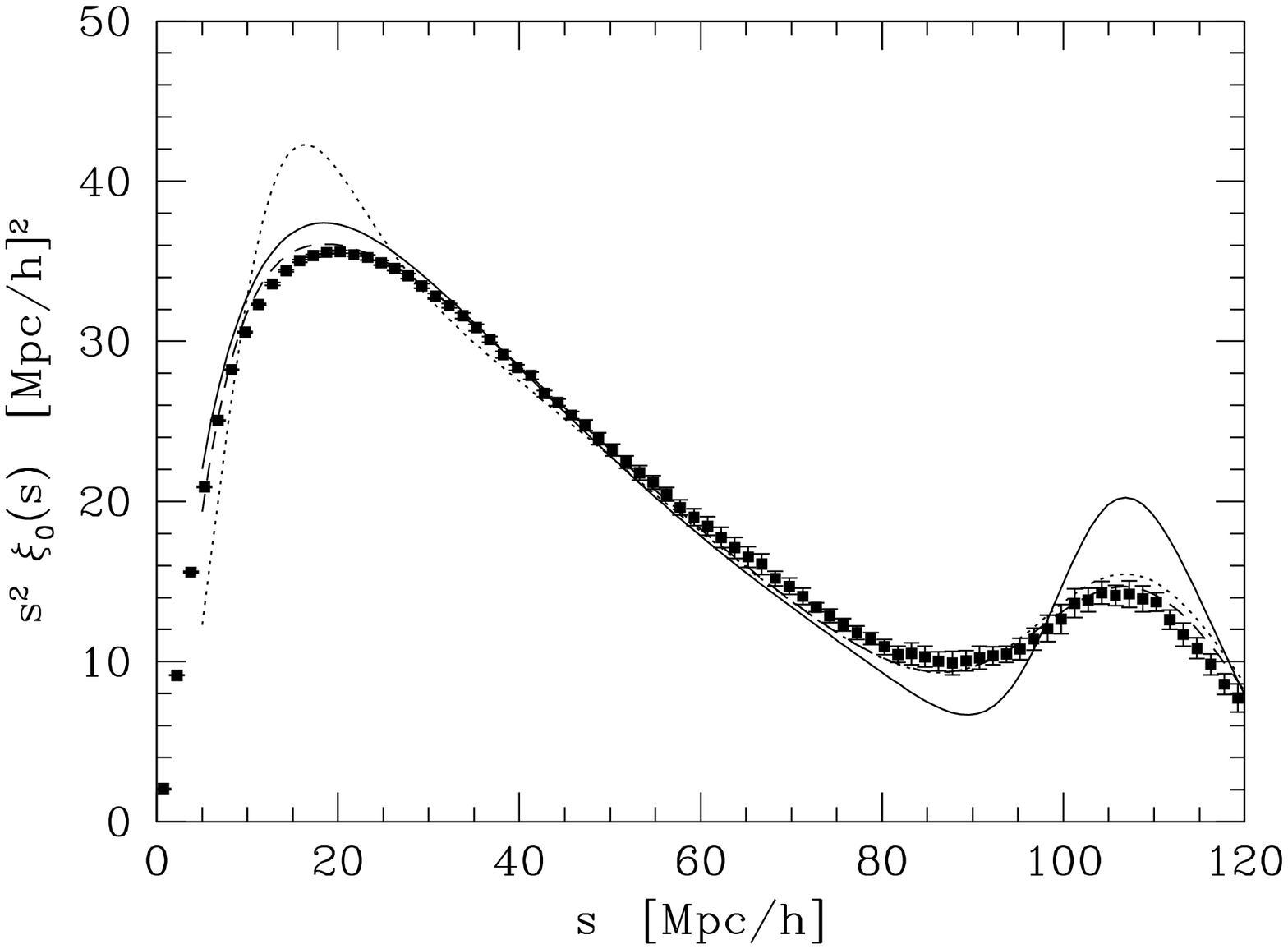}}
    \end{center}
    \caption{(Top) The real-space, matter correlation function, $\xi(r)$, from
    linear theory (solid), LRT (dotted) and CLPT (dashed) compared
    to N-body simulations (squares) at $z=0.55$.
    In order to plot the results with a linear $y$-axis we have multiplied
    $\xi$ by $r^2$, which removes much of the trend from $r\simeq 0-100$Mpc.
    LRT and CLPT agree very well on large scales (the lines can barely be
    distinguished) and agree well with the N-body results.
    LRT overshoots the N-body results below $r\simeq 20\,h^{-1}$Mpc while
    CLPT tracks the N-body results to much smaller scales.
    Linear theory overshoots at $r\simeq 20\,h^{-1}$Mpc and at
    $r\simeq 100\,h^{-1}$Mpc.
    (Bottom) The redshift-space, monopole, matter correlation function,
    $\xi_0(s)$, from linear theory (solid), LRT (dotted) and
    CLPT (dashed) compared to N-body simulations (squares).  The qualitative
    behavior is as for $\xi(r)$.}
    \label{fig:mono_matt}
\end{figure}

\section{Results} \label{sec:results}

Having presented the formalism and rationale behind our resummation,
we now compare the results of our ``convolution Lagrangian Perturbation
Theory'' (CLPT) to linear theory and to the earlier work of
\citet{Mat08a,Mat08b}.  This is the most natural comparison, since our
work is largely an extension of LRT and a partial resummation of that
formalism.

Fig.~\ref{fig:mono_matt} shows the (monopole) matter correlation function
in real- and redshift-space.
The solid line shows linear theory while the dashed and dotted lines show
our CLPT and Matsubara's LRT respectively.
In redshift-space we have used the formalism of \citet{Kai87} as our
``linear theory''.
The points are from the N-body simulations described in \citet{ReiWhi11,Whi11}.
Throughout this paper we compare exclusively with $z=0.55$ simulation outputs.
Note that linear theory provides a poor approximation near the peak of
the correlation function (at $100\,h^{-1}$Mpc) in both real- and redshift-space,
as is well known and we have discussed previously.
On large scales LRT and CLPT are nearly indistinguishable, as expected.
However on smaller scales the resummation inherent in our approach allows
CLPT to track the N-body results to smaller scales than LRT.

\begin{figure}
    \begin{center}
    \resizebox{3in}{!}{\includegraphics{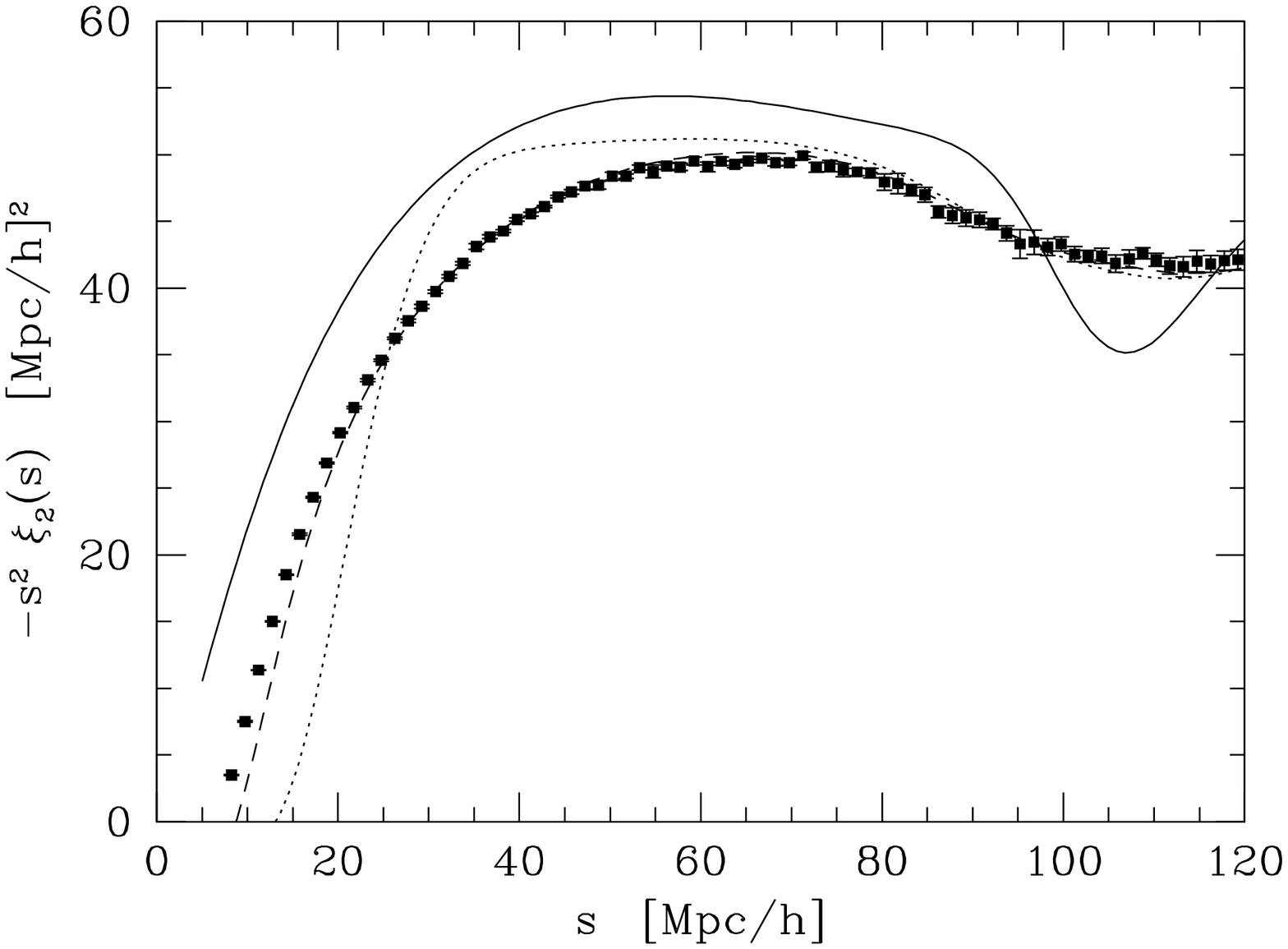}}
    \resizebox{3in}{!}{\includegraphics{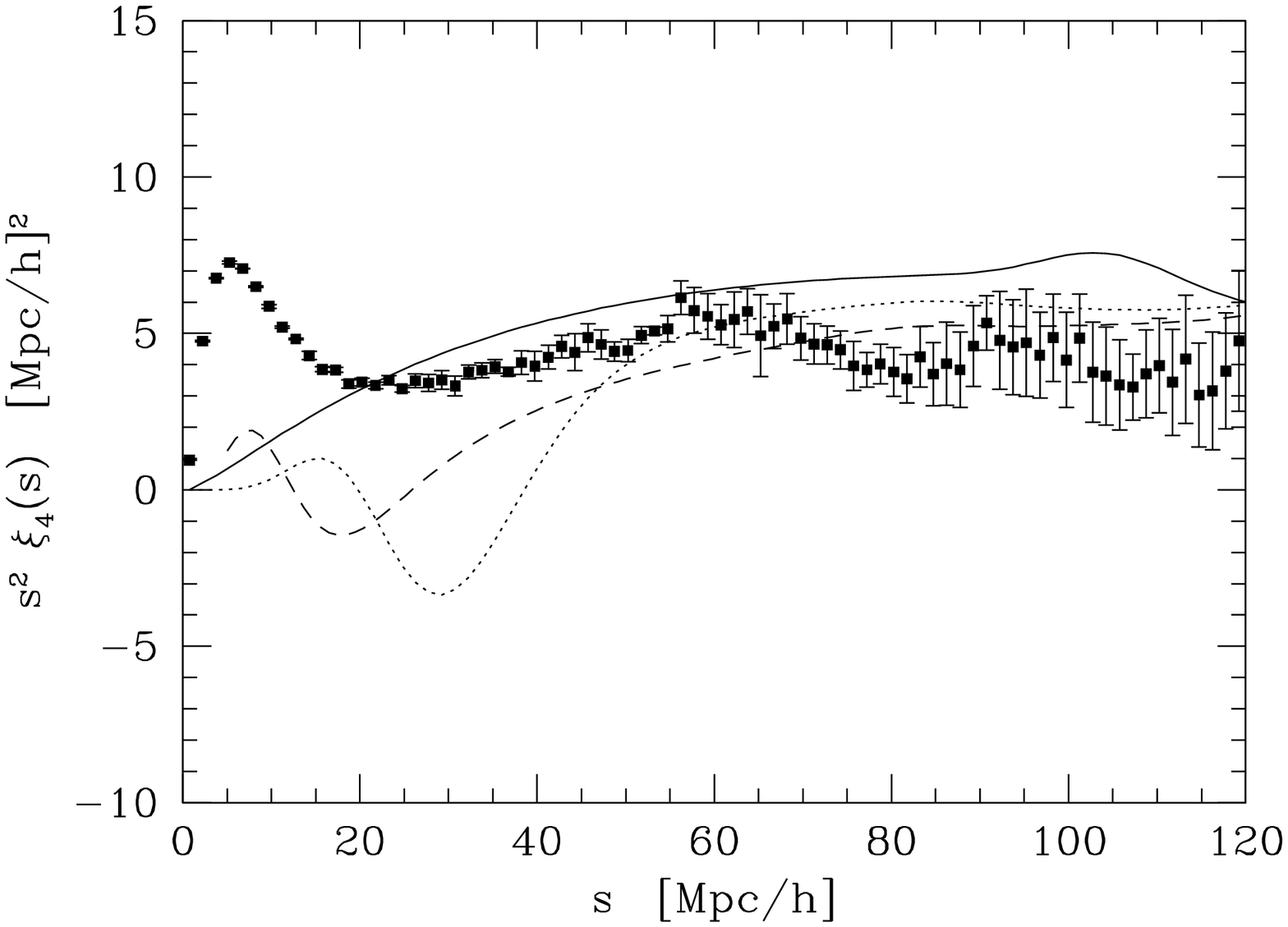}}
    \end{center}
    \caption{The redshift-space, quadrupole and hexadecapole, matter
    correlation functions, $\xi_2(s)$ and $\xi_4(s)$, from linear theory
    (solid), LRT (dotted) and CLPT (dashed) compared to N-body
    simulations (squares) at $z=0.55$.
    For the quadrupole LRT and CLPT agree very well on large scales (and
    agree well with the N-body results) but LRT departs from the N-body
    results at much larger scales.  For the hexadecapole the disagreement
    between N-body, CLPT, LRT and linear theory breaks down at larger
    scales than for the quadrupole.}
    \label{fig:quad_matt}
\end{figure}

The comparison with the quadrupole and hexadecapole moments of the
redshift-space correlation function is very similar (Fig.~\ref{fig:quad_matt}).
Both LRT and CLPT provide a better fit than linear theory to the quadrupole
and hexadecapole moments at large scales, but all theories depart from the
N-body results at larger scales than for the monopole.  The level of agreement
is worse for the hexadecapole, but that moment is also quite small.

\begin{figure}
    \begin{center}
    \resizebox{3in}{!}{\includegraphics{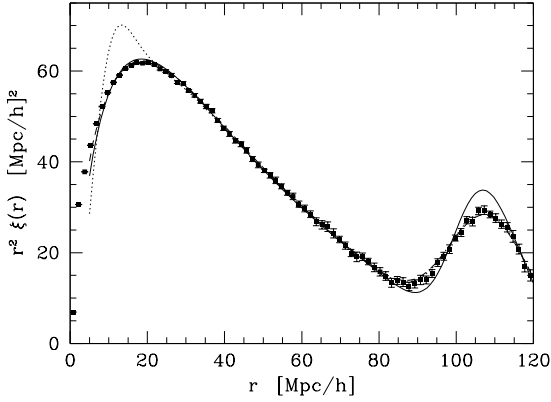}}
    \end{center}
    \caption{The real-space, correlation function for halos with
    $12.8<{\rm lg}M_h<13.1$ computed in linear theory (solid),
    LRT (dotted) and CLPT (dashed) compared to N-body simulations (squares) at
    $z=0.55$.
     In this plot we allowed $\Fp$ and $\Fpp$ to vary independently to obtain the
     best agreement with the N-body results.}
    \label{fig:halo_real}
\end{figure}

\begin{figure}
    \begin{center}
    \resizebox{3in}{!}{\includegraphics{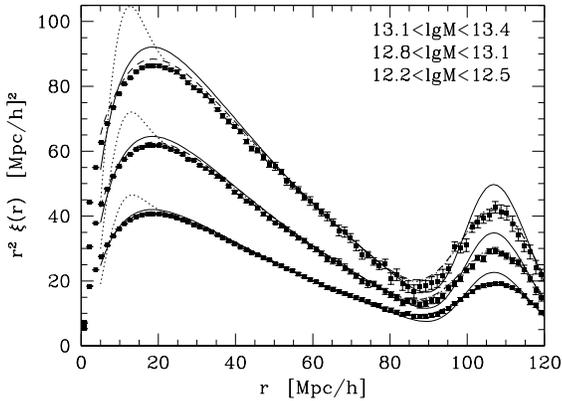}}
    \end{center}
    \caption{The real-space, correlation function for halos in three
    mass bins computed in linear theory (solid), LRT (dotted) and
    CLPT (dashed) compared to N-body simulations (squares) at $z=0.55$ for three
    different mass ranges each a factor of two in width: from bottom
    to top $12.2<{\rm lg}M_h<12.5$, $12.8<{\rm lg}M_h<13.1$ and
    $13.1<{\rm lg}M_h<13.4$ with masses in $h^{-1}M_\odot$.
    In this plot we enforced the peak-background split relation to determine
    $\Fpp$ in terms of the best fit $\Fp$, i.e.~the theory has only one
    free parameter.}
    \label{fig:halo_many}
\end{figure}

\begin{figure}
    \begin{center}
    \resizebox{3in}{!}{\includegraphics{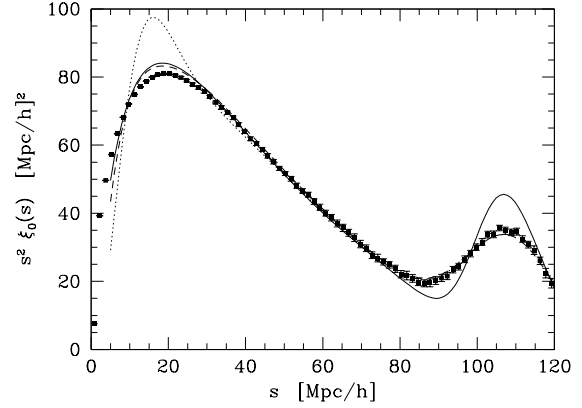}}
    \resizebox{3in}{!}{\includegraphics{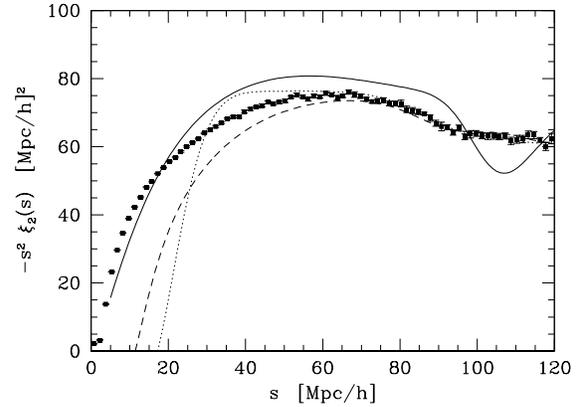}}
    \end{center}
    \caption{The redshift-space, monopole and quadrupole, correlation
    functions for halos computed in linear theory (solid), LRT
    (dotted) and CLPT (dashed) compared to N-body simulations (squares) at
    $z=0.55$.}
    \label{fig:halo_red}
\end{figure}

Fig.~\ref{fig:halo_real} compares the theories for biased tracers, in
this case for halos in the range $12.8<{\rm lg}M_h<13.1$ at $z\simeq 0.55$
though other results are  qualitatively similar (see Fig.~\ref{fig:halo_many}).
The situation is similar to that for the matter: linear theory provides a
poor approximation at large scales, missing the smearing of the acoustic
peak due to the motion of material.  LRT tends to overshoot the N-body
results at small scales, while CLPT provides a good match down to
$\mathcal{O}(10\,h^{-1}{\rm Mpc})$.
Note that we considered two distinct sets of biasing parameters.  In
Figs.~\ref{fig:halo_real} and \ref{fig:halo_red} we allowed the ``renormalized''
bias parameters $\Fp$ and $\Fpp$ to be adjusted independently, while in
Fig.~\ref{fig:halo_many}, we related the two using the peak-background split, as
in \citet{Mat08b,Mat08c}.

Finally we compare the monopole and quadrupole moments of the  redshift
space correlation function of halos to the predictions of CLPT in
Fig.~\ref{fig:halo_red}.  The prediction of the monopole moment is in
relatively good agreement with the N-body measurements, though the level
of agreement at $\sim 20\,h^{-1}$Mpc is clearly not as good as it was with
the matter.  The prediction for the quadrupole is much worse than it was
for the matter.

On large scales the prediction for the quadrupole is dominated by the
same terms as the matter and the term scaling as $\langle F'\rangle$.
The CLPT prediction does not have as much power on small scales as the
N-body results, which have more small-scale power compared to the large-scale
power than was the case for the matter.  The shortfall in power is shared
by the terms which survive when $\langle F'\rangle=0$ and by the terms which
scale as $\langle F'\rangle$.
The failure of our model to match the quadrupole moment on small and
intermediate scales may be due to our assumption of local Lagrangian bias.
While this approximation has received some support from N-body simulations
\citep{RotPor11,Bal12,Chan12,WanSza12} we also expect that terms involving
e.g.~the tidal tensor, can become important for high mass halos \citep{SheChaSco12}. 
Such terms are naturally quadrupolar in nature and may affect the predictions.

\section{Discussion and conclusions}
\label{sec:discuss}

We have presented a new formulation of Lagrangian perturbation theory which
allows accurate predictions of the low-multipole, real- and redshift-space
correlation functions of the mass field and dark matter halos.
Our formulation, which we refer to as ``convolution Lagrangian perturbation
theory'' or CLPT involves a non-perturbative resummation and indeed can
be viewed as a partial resummation of the formalism of
\citet{Mat08a,Mat08b} upon which we have relied heavily.

A key difference between CLPT and LRT or RPT is that we naturally recover
the Zel'dovich approximation as the lowest order of our expansion for the
matter correlation function.
\citet{TasZal12a} have recently emphasized the importance of not splitting
the effects of bulk flows across orders in perturbation theory, and we find
that CLPT (which does not make such a split) does indeed provide better
agreement with N-body results at small scales than LRT (which does).

CLPT works best for the real-space clustering of the matter and halos and
for the monopole of the redshift-space correlation functions.
While the N-body results for the quadrupole and hexadecapole moments of
the redshift-space correlation function for the matter is relatively well
reproduced by CLPT, those moments for the halo correlation function differ
significantly from the CLPT prediction.
We suspect that this difference is due to a limitation in our bias
prescription, in particular that our assumption of local Lagrangian bias
for halos is not sufficiently accurate.
Further work along these lines is clearly warranted.

One possible extension of this work is to use the real-space correlation
function from CLPT in the Gaussian streaming model ansatz of \citet{ReiWhi11}
with $v_{12}$ and $\sigma_{12}$ terms calibrated from N-body simulations or
computed within the context of LPT.  These terms can be computed in our
formalism by generalizing our function $K$ (Eq.~\ref{eq:K}) to include a
$\dot{\Delta}$ contribution and taking functional derivatives of $K$.
We leave this for future work.

Finally, we note that our work may be relevant for efforts to model the
bispectrum within the Lagrangian framework \citep[e.g.,][]{Ram12c}.
\vspace{0.2in}

J.C. and M.W. are supported by the NSF and NASA.
B.A.R. is supported by NASA.  through Hubble Fellowship grant 51280 
awarded by the Space Telescope Science Institute, which is operated 
by the Association of Universities for Research in Astronomy, Inc., 
for NASA, under contract NAS 5-26555.
This work made extensive use of the NASA Astrophysics Data System and
of the {\tt astro-ph} preprint archive at {\tt arXiv.org}.  The
analysis made use of the computing resources of the National Energy
Research Scientific Computing Center.


\newcommand{\aj}{AJ}
\newcommand{\apj}{ApJ}
\newcommand{\apjs}{ApJ Suppl.}
\newcommand{\mnras}{MNRAS}
\newcommand{\araa}{ARA{\&}A}
\newcommand{\aap}{A{\&}A}
\newcommand{\pre}{PRE}
\newcommand{\prd}{Phys. Rev. D}
\newcommand{\apjl}{ApJL}
\newcommand{\physrep}{Physics Reports}
\newcommand{\nat}{Nature}

\begin{appendix}

\section{Bias} \label{sec:bias}

As in \citet{Mat08b}, we note the identity
\begin{equation}
    \int \frac{d\lam}{2\pi}~ \tilde{F}(\lam) e^{-\frac{1}{2} \lam^2 \sigma_R^2} (i\lam)^n = \left\langle F^{(n)} \right\rangle ,
\end{equation}
where $\left\langle F^{(n)} \right\rangle$ is the expectation value of the
$n$th derivative of the Lagrangian bias function, what are referred to as
``renormalized'' bias coefficients $c_n$ in Matsubara 2011.  The mapping
$K \to L$ is therefore achieved by replacing
\begin{align}
    (\lam_1 + \lam_2) &\to -2i \Fp , \\
    \lam_1 \lam_2 &\to -\Fp^2 , \\
    \lam_1^2 \lam_2^2 &\to \Fpp^2 , \\
    (\lam_1 + \lam_2)^2 &\to -2 \left[\Fpp + \Fp^2\right] , \\
    \lam_1 \lam_2 (\lam_1 + \lam_2) &\to 2i \Fp \Fpp , \\
    \lam_1^2 + \lam_2^2 &\to -2 \Fpp .
\end{align}

\section{Lagrangian correlators} \label{sec:correlators}

In this appendix we collect the relevant facts and formulas concerning
Lagrangian correlators that we need for our one-loop theory.  The correlators
are defined by
\begin{equation}
    C_{i_1\cdots i_r}^{mn}(\vq) = \left\langle \delta_1^m \delta_2^n \Delta_{i_1} \cdots \Delta_{i_r} \right\rangle_c ,
\end{equation}
where $\delta_1 = \delta_L(\vq_1)$, $\delta_2 = \delta_L(\vq_2)$, and
$\Delta_i = \Psi_i(\vq_2) - \Psi_i(\vq_1)$.  The subscripted $c$ refers to a
connected moment; since these fields have zero mean, the connected moments
coincide with normal expectation values for orders $m+n+r \le 3$.  

\subsection{Index structure}

By translational symmetry, a Lagrangian correlator can only depend on the
Lagrangian separation $\vq = \vq_2 - \vq_1$.  This imposes strong constraints
on its index structure.  We classify a correlator by its tensor rank, i.e. by
the number of vector indices it carries.  In the following we let $U_i$,
$A_{ij}$, and $W_{ijk}$ denote generic correlators of ranks 1, 2, and 3,
respectively.

Rank-1 correlators must be of the form
\begin{equation}
    U_i(\vq) = U(q)~ \hq_i
\end{equation}
for some scalar function $U(q)$, since (trivially) the only vector quantity
that can be formed from the vector $\vq$ is proportional to $\vq$.  Rank-2
correlators must involve only rotationally invariant rank-2 tensors that can be
formed from the vector $\vq$, i.e. $\delta_{ij}$ or $\hq_i \hq_j$.  Thus their
general form is
\begin{equation}
    A_{ij}(\vq) = X(q)~ \delta_{ij} + Y(q)~ \hq_i \hq_j .
\end{equation}
Likewise, rank-3 correlators are of the form
\begin{equation}
  W_{ijk}(\vq) = V_1(q)~ \hq_i \delta_{jk} + V_2(q)~ \hq_j \delta_{ki}
  + V_3(q)~ \hq_k \delta_{ij} + T(q)~ \hq_i \hq_j \hq_k .
\end{equation}
We remind the reader that we adopt the shorthand
\begin{equation}
    U_{i}^{10} \to U_i , \quad
    A_{ij}^{00} \to A_{ij} , \quad \text{and }
    W_{ijk}^{00} \to W_{ijk} ,
\end{equation}
since these combinations arise frequently.

In general, correlators of even rank are even functions of $\vq$, while those
of odd rank are odd.  This implies that the correlator $C_{i_1\dots i_r}^{mn}$
is symmetric in the indices $m$ and $n$, as the following chain of equalities
shows:
\begin{equation}
  \begin{aligned}
    C_{i_1\dots i_r}^{mn}(\vq) &=
    \left\langle \delta_1^m \delta_2^n \Delta_{i_1} \cdots
    \Delta_{i_r} \right\rangle_c \\
    &= \left\langle \delta_L(\vq_1)^m \delta_L(\vq_2)^n
    [\Psi_{i_1}(\vq_2)-\Psi_{i_1}(\vq_1)] \cdots \right. \\
    & \left. \cdots
    [\Psi_{i_r}(\vq_2)-\Psi_{i_r}(\vq_1)] \right\rangle_c \\
    &= (-1)^r \left\langle \delta_L(\vq_2)^n \delta_L(\vq_1)^m
    [\Psi_{i_1}(\vq_1)-\Psi_{i_1}(\vq_2)] \cdots \right. \\
    & \cdots \left.
    [\Psi_{i_r}(\vq_1)-\Psi_{i_r}(\vq_2)] \right\rangle_c \\
    &= (-1)^r C_{i_1\dots i_r}^{nm}(-\vq) \\
    &= C_{i_1\dots i_r}^{nm}(\vq) .
  \end{aligned}
\end{equation}

We can solve for the coefficients in these expansions by contracting
against tensors and solving the resulting simultaneous equations, e.g.
for the components of $W_{ijk}$:
\begin{equation}
    \begin{aligned}
    3V_1 + V_2 + V_3 + T &= W_{ijk}~ \hq_i \delta_{jk} , \\
    V_1 + 3V_2 + V_3 + T &= W_{ijk}~ \hq_j \delta_{ki} , \\
    V_1 + V_2 + 3V_3 + T &= W_{ijk}~ \hq_k \delta_{ij} , \\
    V_1 + V_2 + V_3 + T &= W_{ijk}~ \hq_i \hq_j \hq_k .
    \end{aligned}
\end{equation}

\subsection{Perturbative orders}

The LPT expansion of the field $\vDelta$ has the form
\begin{equation}
    \vDelta = \vDelta^{(1)} + \vDelta^{(2)} + \vDelta^{(3)} + \cdots ,
\end{equation}
where $\vDelta^{(a)}$ involves $a$ factors of the linear density field
$\delta_L$.  The correlators $C^{mn}_{i_1\cdots i_r}$ may therefore be expanded
as
\begin{equation}
    C^{mn}_{i_1\cdots i_r} = \sum_{a_1=1}^\infty \cdots \sum_{a_r=1}^\infty C^{mn(a_1\cdots a_r)}_{i_1\cdots i_r} ,
\end{equation}
where
$C^{mn(a_1\cdots a_r)}_{i_1\cdots i_r} = \left\langle \delta_1^m \delta_2^n \Delta_{i_1}^{(a_1)} \cdots \Delta_{i_r}^{(a_r)} \right\rangle_c$.
Since $\delta_L$ is Gaussian, many of these terms vanish.  Here we display the
breakdown for each of the quantities introduced in Section \ref{sec:oneloop},
up to order $O(P_L^2)$:
\begin{align}
  U_i &= U_i^{(1)} + U_i^{(3)} + \cdots , \\
  A_{ij} &= A_{ij}^{(11)} + A_{ij}^{(22)} + A_{ij}^{(13)} + A_{ij}^{(31)}
          + \cdots , \\
  W_{ijk} &= W_{ijk}^{(112)} + W_{ijk}^{(121)} + W_{ijk}^{(211)} + \cdots , \\
  U_i^{20} &= U_i^{20(2)} + \cdots , \\
  U_i^{11} &= U_i^{11(2)} + \cdots , \\
  A_{ij}^{10} &= A_{ij}^{10(12)} + A_{ij}^{10(21)} + \cdots ,
\end{align}

\subsection{Scalar components}

Given the index structure described in the previous subsection, evaluating the
Lagrangian correlators reduces to computing a set of scalar functions of $q$.
In order to maintain notational consistency with \citet{Mat08b} we make use of
his definitions of $Q$ and $R$.
\begin{equation}
  R_n(k) = \frac{k^3}{4\pi^2} P_L(k)\int_0^\infty dr\ P_L(kr)\widetilde{R}_n(r)
\end{equation}
and
\begin{equation}
  Q_{n}(k) = \frac{k^{3}}{4\pi^2} \int_{0}^{\infty}dr\,P_{L}(kr)
  \int_{-1}^{1}dx\,P_{L}(k\sqrt{y})Q_{n}(r,x) \,\,,
\label{eq:qndef}
\end{equation}
where $y(r,x) = 1 + r^{2} - 2rx$ and the $Q_{n}$ are given by
$$
\begin{array}{ll}
\displaystyle Q_{1} = \frac{r^2(1-x^2)^2}{y^2}, &\displaystyle  Q_{2} = \frac{(1-x^2)rx(1-rx)}{y^2}, \\
\displaystyle Q_{3} = \frac{x^2(1-rx)^2}{y^2}, &\displaystyle   Q_{4} = \frac{1-x^2}{y^2}, \\
\displaystyle Q_{5} = \frac{rx(1-x^2)}{y}, &\displaystyle   Q_{6} = \frac{(1-3rx)(1-x^2)}{y}, \\
\displaystyle Q_{7} = \frac{x^2 (1-rx)}{y}, &\displaystyle   Q_{8} = \frac{r^2(1-x^2)}{y}, \\
\displaystyle Q_{9} = \frac{rx(1-rx)}{y}, &\displaystyle  Q_{10} = 1-x^2, \\
\multicolumn{2}{l}{\displaystyle Q_{11} = x^2,\,\, \displaystyle  Q_{12}=rx,\,\, \displaystyle Q_{13}=r^2}
\end{array}
$$
and
$$
\begin{array}{lcl}
\widetilde{R}_1(r) &=& \int_{-1}^{+1}dx
  \ \displaystyle\frac{r^2(1-x^2)^2}{1+r^2-2rx} \\
 & & \\
\widetilde{R}_2(r) &=& \int_{-1}^{+1}dx
  \ \displaystyle\frac{(1-x^2)rx(1-rx)}{1+r^2-2rx}
\end{array}
$$
In the following, equation references prefaced with ``M'' indicate equations
in \citet{Mat08b}.

The expression for $A_{ij}=A^{00}_{ij}$ is derived in detail below.  The
other components we need are
\begin{eqnarray}
  A_{ij}^{10}(\vq) &=& \langle \delta_1 \Delta_i \Delta_j \rangle_c \\
  &=& X_{10}(q) \delta_{ij} + Y_{10}(q) \hq_i \hq_j
\end{eqnarray}
with
\begin{align}
    \xi_L(q) &= \frac{1}{2\pi^2} \int_0^\infty dk~ k^2 P_L(k) j_0(kq) \\
    V_1^{(112)}(q) &= \frac{1}{2\pi^2} \int_0^\infty \frac{dk}{k}~
    \left(-\frac{3}{7}\right) R_1 j_1(kq) \\
    V_3^{(112)}(q) &= \frac{1}{2\pi^2} \int_0^\infty \frac{dk}{k}~
    \left(-\frac{3}{7}\right) Q_1 j_1(kq) \\
    S^{(112)}(q) &= \frac{1}{2\pi^2} \int_0^\infty \frac{dk}{k}~
    \frac{3}{7} \left[2 R_1+4 R_2+Q_1+2Q_2\right] \frac{j_2(kq)}{kq} \\
    T^{(112)}(q) &= \frac{1}{2\pi^2} \int_0^\infty \frac{dk}{k}~
    \left(-\frac{3}{7}\right) \times \notag \\
    & \quad \left[2R_1+4R_2+Q_1+2Q_2\right] j_3(kq) \\
    U^{(1)}(q) &= \frac{1}{2\pi^2} \int_0^\infty dk~ k~ (-1) P_L(k) j_1(kq) \\
    U^{(3)}(q) &= \frac{1}{2\pi^2} \int_0^\infty dk~ k~
    \left(-\frac{5}{21}\right) R_1 j_1(kq) \\
    U_{20}^{(2)}(q) &= \frac{1}{2\pi^2} \int_0^\infty dk~ k~
    \left(-\frac{3}{7}\right) Q_8 j_1(kq) \\
    U_{11}^{(2)}(q) &= \frac{1}{2\pi^2} \int_0^\infty dk~ k~
    \left(-\frac{6}{7}\right) [R_1+R_2] j_1(kq) \\
    X_{10}^{(12)}(q) &= \frac{1}{2\pi^2} \int_0^\infty dk~
    \frac{1}{14} \left\{ \vphantom{\int}
    2 [R_1-R_2] + 3R_1 j_0(kq) \right. \notag \\
    &\quad \left. -3 [3R_1+4R_2+2Q_5] \frac{j_1(kq)}{kq} \right\} \\
    Y_{10}^{(12)}(q) &= \frac{1}{2\pi^2} \int_0^\infty dk~
    \left(-\frac{3}{14}\right) [3R_1+4R_2+2Q_5] \times \notag \\
    &\quad \left[j_0(kq) - 3 \frac{j_1(kq)}{kq}\right]
\end{align}
where the arguments of the $R_n$ and $Q_n$ terms are $k$ and have been
omitted for brevity.
The remaining equations, for $X^{(11)}$, $X^{(22)}$, $X^{(13)}$,
$Y^{(11)}$, $Y^{(22)}$, $Y^{(13)}$ are presented and derived in the
next section,

\subsection{Example}

We provide here an example of how to obtain the formulae of the previous
subsection.  We focus on $A_{ij} = \langle \Delta_i \Delta_j \rangle_c$, since
this is the most important of the Lagrangian correlators in our theory.

By the definition of $\vDelta$,
\begin{equation}
  \Delta_i = \Psi_i(\vq_2) - \Psi_i(\vq_1)
  = \int \frac{d^3p}{(2\pi)^3}
  \left(e^{i\vp\cdot\vq_2} - e^{i\vp\cdot\vq_1}\right) \tilde{\Psi}_i(\vp) ,
\end{equation}
and therefore
\begin{eqnarray}
  A_{ij} &=& \int \frac{d^3p_1}{(2\pi)^3} \frac{d^3p_2}{(2\pi)^3}
  \left(e^{i\vp_1\cdot\vq_2} - e^{i\vp_1\cdot\vq_1}\right)
  \left(e^{i\vp_2\cdot\vq_2} - e^{i\vp_2\cdot\vq_1}\right) \nonumber \\
  &\times&
  \left\langle \tilde{\Psi}_i(\vp_1) \tilde{\Psi}_j(\vp_2) \right\rangle_c .
\label{eq:Aij}
\end{eqnarray}
From Eq. (M.A9), the Fourier space 2-point function here is
\begin{equation}
    \label{eq:CM}
    \left\langle \tilde{\Psi}_i(\vp_1) \tilde{\Psi}_j(\vp_2) \right\rangle_c
        = -(2\pi)^3 \delta_D^3(\vp_1 + \vp_2) C_{ij}(\vp_1) .
\end{equation}
The quantity $C_{ij}(\vk)$ here has contributions at both tree and 1-loop
level,
\begin{gather*}
  C_{ij}^{(11)}(\vk) = -\frac{k_i k_j}{k^4} P_L(k) , \tag{M.A52} \\
  C_{ij}^{(22)}(\vk) = -\frac{9}{98} \frac{k_i k_j}{k^4} Q_1(k) , \tag{M.A53} \\
  C_{ij}^{(13)}(\vk) = C_{ij}^{(31)}(\vk) = -\frac{5}{21}
        \frac{k_i k_j}{k^4} R_1(k) . \tag{M.A54}
\end{gather*}
These terms are all of the form $C_{ij} = -(k_i k_j/k^4)a(k)$ for scalar
$a(k)$, as is guaranteed by rotational symmetry.
With the substitution of Eq. (\ref{eq:CM}) into Eq. (\ref{eq:Aij}), we have
\begin{equation}
  A_{ij} = \int \frac{d^3k}{(2\pi)^3}
  \left(2 - e^{i\vk\cdot\vq} - e^{-i\vk\cdot\vq}\right)
  \frac{k_i k_j}{k^4} a(k) .
\end{equation}
Contracting this quantity first by $\delta_{ij}$ and then by $\hq_i \hq_j$, we
obtain the system of equations
\begin{align}
  A_{ij} \delta_{ij} &= 3X + Y = \int \frac{d^3k}{(2\pi)^3}
  \left(2 - e^{i\vk\cdot\vq} - e^{-i\vk\cdot\vq}\right) \frac{1}{k^2} a(k) , \\
  A_{ij} \hq_i \hq_j &= X + Y = \int \frac{d^3k}{(2\pi)^3}
  \left(2 - e^{i\vk\cdot\vq} - e^{-i\vk\cdot\vq}\right)
  \frac{(\hk\cdot\hq)^2}{k^2} a(k) .
\end{align}
Letting $\mu = \hk\cdot\hq$ and using the Bessel function identities in
Appendix \ref{sec:reference} we may perform the angular integrations,
\begin{align}
  3X + Y &= \frac{1}{2\pi^2} \int_0^\infty k^2 dk \frac{1}{2}
  \int_{-1}^1 d\mu~ \left(2 - e^{ikq\mu} - e^{-ikq\mu}\right)
  \frac{1}{k^2} a(k) \notag \\
  &= \frac{1}{2\pi^2} \int_0^\infty dk \left[2 - 2 j_0(kq)\right] a(k) , \\
  X + Y &= \frac{1}{2\pi^2} \int_0^\infty k^2 dk \frac{1}{2}
  \int_{-1}^1 d\mu~ \left(2 - e^{ikq\mu} - e^{-ikq\mu}\right)
  \frac{\mu^2}{k^2} a(k) \notag \\
  &= \frac{1}{2\pi^2} \int_0^\infty dk
  \left[\frac{2}{3} - 2 j_0(kq) + 4 \frac{j_1(kq)}{kq}\right] a(k) ,
\end{align}
from which we obtain
\begin{align}
  X(q) &= \frac{1}{2\pi^2} \int_0^\infty dk~ a(k)
  \left[\frac{2}{3} - 2 \frac{j_1(kq)}{kq}\right] , \\
  Y(q) &= \frac{1}{2\pi^2} \int_0^\infty dk~ a(k)
  \left[-2 j_0(kq) + 6 \frac{j_1(kq)}{kq}\right] .
\end{align}
Explicitly, up to 1-loop order, the contributions to $X(q)$ and $Y(q)$ are
\begin{align}
  X^{(11)}(q) &= \frac{1}{2\pi^2} \int_0^\infty dk
  \ P_L(k) \left[\frac{2}{3} - 2 \frac{j_1(kq)}{kq}\right] , \\
  X^{(22)}(q) &= \frac{1}{2\pi^2} \int_0^\infty dk
  \ \frac{9}{98} Q_1(k) \left[\frac{2}{3} - 2 \frac{j_1(kq)}{kq}\right] , \\
  X^{(13)}(q) &= \frac{1}{2\pi^2} \int_0^\infty dk
  \ \frac{5}{21} R_1(k) \left[\frac{2}{3} - 2 \frac{j_1(kq)}{kq}\right] , \\
  Y^{(11)}(q) &= \frac{1}{2\pi^2} \int_0^\infty dk
  \ P_L(k) \left[-2 j_0(kq) + 6 \frac{j_1(kq)}{kq}\right] , \\
  Y^{(22)}(q) &= \frac{1}{2\pi^2} \int_0^\infty dk
  \ \frac{9}{98} Q_1(k) \left[-2 j_0(kq) + 6 \frac{j_1(kq)}{kq}\right] , \\
  Y^{(13)}(q) &= \frac{1}{2\pi^2} \int_0^\infty dk
  \ \frac{5}{21} R_1(k) \left[-2 j_0(kq) + 6 \frac{j_1(kq)}{kq}\right] .
\end{align}
Note that each of these quantities approaches 0 as $q \to 0$.

\section{Reference formulae} \label{sec:reference}

\subsection{Gaussian integrals}

In our theory we make use of the basic Gaussian integral
\begin{equation}
    Q(\vb) \equiv \int \frac{d^3k}{(2\pi)^3}~ e^{-\frac{1}{2} \vk^T \mat{A} \vk + i \vb\cdot\vk}
        = \frac{1}{(2\pi)^{3/2} |A|^{1/2}} e^{-\frac{1}{2} \vb^T \mat{A}^{-1} \vb} .
\end{equation}
where $|A|$ denotes the determinant of the $3\times3$ matrix $\mat{A}$.  By
successive applications of the operator $-i\partial/\partial b_i$,
we also have
\begin{eqnarray}
  \int \frac{d^3k}{(2\pi)^3}\ G(\mathbf{k}) k_i
  &=& i (\mat{A}^{-1} \vb)_i Q(\vb) \quad , \\
  \int \frac{d^3k}{(2\pi)^3}\ G(\mathbf{k}) k_i k_j
  &=& \left[(\mat{A}^{-1})_{ij} - (\mat{A}^{-1} \vb)_i
    (\mat{A}^{-1} \vb)_j\right] Q(\vb) , \\
  \int \frac{d^3k}{(2\pi)^3}\ G(\mathbf{k}) k_i k_j k_k
  &=& i \Big[ (\mat{A}^{-1})_{ij} (\mat{A}^{-1} \vb)_k
   +          (\mat{A}^{-1})_{ki} (\mat{A}^{-1} \vb)_j \nonumber \\
  &+& (\mat{A}^{-1})_{jk} (\mat{A}^{-1} \vb)_i \nonumber \\
  &-& (\mat{A}^{-1} \vb)_i (\mat{A}^{-1} \vb)_j (\mat{A}^{-1} \vb)_k \Big]
  Q(\vb) .
\end{eqnarray}
where we have written
\begin{equation}
  G(\mathbf{k}) = e^{-\frac{1}{2} \vk^T \mat{A} \vk + i \vb\cdot\vk}
\end{equation}
for notational compactness.

\subsection{Spherical Bessel functions}

In performing the integrals in the previous sections we have found the
following spherical Bessel function identities and integrals to be useful:
\begin{align}
    j_{n-1}(x) + j_{n+1}(x) &= (2n+1) \frac{j_n(x)}{x} \\
    n j_{n-1}(x) - (n+1) j_{n+1}(x) &= (2n+1) \frac{d}{dx} j_n(x) \\
    \frac{1}{2} \int_{-1}^1 d\mu\ e^{ix\mu} &= j_0(x) \\
    \frac{1}{2} \int_{-1}^1 d\mu\ \mu\, e^{ix\mu} &= i j_1(x) \\
    \frac{1}{2} \int_{-1}^1 d\mu\ \mu^2\, e^{ix\mu} &=
         \frac{1}{3} j_0(x) - \frac{2}{3} j_2(x) \\
    &= j_0(x) - 2 \frac{j_1(x)}{x} \\
    \frac{1}{2} \int_{-1}^1 d\mu\ \mu^3\, e^{ix\mu} &=
    i \left[\frac{3}{5} j_1(x) - \frac{2}{5} j_3(x)\right]
\end{align}

\end{appendix}

\label{lastpage}
\end{document}